\documentclass[twocolumn]{aastex631}
\pdfoutput=1

\begin{document}


\title{A Search for Coincident Neutrino Emission from Fast Radio Bursts with Seven Years of IceCube Cascade Events}

\affiliation{III. Physikalisches Institut, RWTH Aachen University, D-52056 Aachen, Germany}
\affiliation{Department of Physics, University of Adelaide, Adelaide, 5005, Australia}
\affiliation{Dept. of Physics and Astronomy, University of Alaska Anchorage, 3211 Providence Dr., Anchorage, AK 99508, USA}
\affiliation{Dept. of Physics, University of Texas at Arlington, 502 Yates St., Science Hall Rm 108, Box 19059, Arlington, TX 76019, USA}
\affiliation{CTSPS, Clark-Atlanta University, Atlanta, GA 30314, USA}
\affiliation{School of Physics and Center for Relativistic Astrophysics, Georgia Institute of Technology, Atlanta, GA 30332, USA}
\affiliation{Dept. of Physics, Southern University, Baton Rouge, LA 70813, USA}
\affiliation{Dept. of Physics, University of California, Berkeley, CA 94720, USA}
\affiliation{Lawrence Berkeley National Laboratory, Berkeley, CA 94720, USA}
\affiliation{Institut f{\"u}r Physik, Humboldt-Universit{\"a}t zu Berlin, D-12489 Berlin, Germany}
\affiliation{Fakult{\"a}t f{\"u}r Physik {\&} Astronomie, Ruhr-Universit{\"a}t Bochum, D-44780 Bochum, Germany}
\affiliation{Universit{\'e} Libre de Bruxelles, Science Faculty CP230, B-1050 Brussels, Belgium}
\affiliation{Vrije Universiteit Brussel (VUB), Dienst ELEM, B-1050 Brussels, Belgium}
\affiliation{Department of Physics and Laboratory for Particle Physics and Cosmology, Harvard University, Cambridge, MA 02138, USA}
\affiliation{Dept. of Physics, Massachusetts Institute of Technology, Cambridge, MA 02139, USA}
\affiliation{Dept. of Physics and The International Center for Hadron Astrophysics, Chiba University, Chiba 263-8522, Japan}
\affiliation{Department of Physics, Loyola University Chicago, Chicago, IL 60660, USA}
\affiliation{Dept. of Physics and Astronomy, University of Canterbury, Private Bag 4800, Christchurch, New Zealand}
\affiliation{Dept. of Physics, University of Maryland, College Park, MD 20742, USA}
\affiliation{Dept. of Astronomy, Ohio State University, Columbus, OH 43210, USA}
\affiliation{Dept. of Physics and Center for Cosmology and Astro-Particle Physics, Ohio State University, Columbus, OH 43210, USA}
\affiliation{Niels Bohr Institute, University of Copenhagen, DK-2100 Copenhagen, Denmark}
\affiliation{Dept. of Physics, TU Dortmund University, D-44221 Dortmund, Germany}
\affiliation{Dept. of Physics and Astronomy, Michigan State University, East Lansing, MI 48824, USA}
\affiliation{Dept. of Physics, University of Alberta, Edmonton, Alberta, Canada T6G 2E1}
\affiliation{Erlangen Centre for Astroparticle Physics, Friedrich-Alexander-Universit{\"a}t Erlangen-N{\"u}rnberg, D-91058 Erlangen, Germany}
\affiliation{Physik-department, Technische Universit{\"a}t M{\"u}nchen, D-85748 Garching, Germany}
\affiliation{D{\'e}partement de physique nucl{\'e}aire et corpusculaire, Universit{\'e} de Gen{\`e}ve, CH-1211 Gen{\`e}ve, Switzerland}
\affiliation{Dept. of Physics and Astronomy, University of Gent, B-9000 Gent, Belgium}
\affiliation{Dept. of Physics and Astronomy, University of California, Irvine, CA 92697, USA}
\affiliation{Karlsruhe Institute of Technology, Institute for Astroparticle Physics, D-76021 Karlsruhe, Germany }
\affiliation{Karlsruhe Institute of Technology, Institute of Experimental Particle Physics, D-76021 Karlsruhe, Germany }
\affiliation{Dept. of Physics, Engineering Physics, and Astronomy, Queen's University, Kingston, ON K7L 3N6, Canada}
\affiliation{Department of Physics {\&} Astronomy, University of Nevada, Las Vegas, NV, 89154, USA}
\affiliation{Nevada Center for Astrophysics, University of Nevada, Las Vegas, NV 89154, USA}
\affiliation{Dept. of Physics and Astronomy, University of Kansas, Lawrence, KS 66045, USA}
\affiliation{Department of Physics and Astronomy, UCLA, Los Angeles, CA 90095, USA}
\affiliation{Centre for Cosmology, Particle Physics and Phenomenology - CP3, Universit{\'e} catholique de Louvain, Louvain-la-Neuve, Belgium}
\affiliation{Department of Physics, Mercer University, Macon, GA 31207-0001, USA}
\affiliation{Dept. of Astronomy, University of Wisconsin{\textendash}Madison, Madison, WI 53706, USA}
\affiliation{Dept. of Physics and Wisconsin IceCube Particle Astrophysics Center, University of Wisconsin{\textendash}Madison, Madison, WI 53706, USA}
\affiliation{Institute of Physics, University of Mainz, Staudinger Weg 7, D-55099 Mainz, Germany}
\affiliation{Department of Physics, Marquette University, Milwaukee, WI, 53201, USA}
\affiliation{Institut f{\"u}r Kernphysik, Westf{\"a}lische Wilhelms-Universit{\"a}t M{\"u}nster, D-48149 M{\"u}nster, Germany}
\affiliation{Bartol Research Institute and Dept. of Physics and Astronomy, University of Delaware, Newark, DE 19716, USA}
\affiliation{Dept. of Physics, Yale University, New Haven, CT 06520, USA}
\affiliation{Columbia Astrophysics and Nevis Laboratories, Columbia University, New York, NY 10027, USA}
\affiliation{Dept. of Physics, University of Oxford, Parks Road, Oxford OX1 3PU, UK}
\affiliation{Dipartimento di Fisica e Astronomia Galileo Galilei, Universit{\`a} Degli Studi di Padova, 35122 Padova PD, Italy}
\affiliation{Dept. of Physics, Drexel University, 3141 Chestnut Street, Philadelphia, PA 19104, USA}
\affiliation{Physics Department, South Dakota School of Mines and Technology, Rapid City, SD 57701, USA}
\affiliation{Dept. of Physics, University of Wisconsin, River Falls, WI 54022, USA}
\affiliation{Dept. of Physics and Astronomy, University of Rochester, Rochester, NY 14627, USA}
\affiliation{Department of Physics and Astronomy, University of Utah, Salt Lake City, UT 84112, USA}
\affiliation{Oskar Klein Centre and Dept. of Physics, Stockholm University, SE-10691 Stockholm, Sweden}
\affiliation{Dept. of Physics and Astronomy, Stony Brook University, Stony Brook, NY 11794-3800, USA}
\affiliation{Dept. of Physics, Sungkyunkwan University, Suwon 16419, Korea}
\affiliation{Institute of Physics, Academia Sinica, Taipei, 11529, Taiwan}
\affiliation{Dept. of Physics and Astronomy, University of Alabama, Tuscaloosa, AL 35487, USA}
\affiliation{Dept. of Astronomy and Astrophysics, Pennsylvania State University, University Park, PA 16802, USA}
\affiliation{Dept. of Physics, Pennsylvania State University, University Park, PA 16802, USA}
\affiliation{Dept. of Physics and Astronomy, Uppsala University, Box 516, S-75120 Uppsala, Sweden}
\affiliation{Dept. of Physics, University of Wuppertal, D-42119 Wuppertal, Germany}
\affiliation{Deutsches Elektronen-Synchrotron DESY, Platanenallee 6, 15738 Zeuthen, Germany }

\author[0000-0001-6141-4205]{R. Abbasi}
\affiliation{Department of Physics, Loyola University Chicago, Chicago, IL 60660, USA}

\author[0000-0001-8952-588X]{M. Ackermann}
\affiliation{Deutsches Elektronen-Synchrotron DESY, Platanenallee 6, 15738 Zeuthen, Germany }

\author{J. Adams}
\affiliation{Dept. of Physics and Astronomy, University of Canterbury, Private Bag 4800, Christchurch, New Zealand}

\author{N. Aggarwal}
\affiliation{Dept. of Physics, University of Alberta, Edmonton, Alberta, Canada T6G 2E1}

\author[0000-0003-2252-9514]{J. A. Aguilar}
\affiliation{Universit{\'e} Libre de Bruxelles, Science Faculty CP230, B-1050 Brussels, Belgium}

\author[0000-0003-0709-5631]{M. Ahlers}
\affiliation{Niels Bohr Institute, University of Copenhagen, DK-2100 Copenhagen, Denmark}

\author[0000-0002-9534-9189]{J.M. Alameddine}
\affiliation{Dept. of Physics, TU Dortmund University, D-44221 Dortmund, Germany}

\author{A. A. Alves Jr.}
\affiliation{Karlsruhe Institute of Technology, Institute for Astroparticle Physics, D-76021 Karlsruhe, Germany }

\author{N. M. Amin}
\affiliation{Bartol Research Institute and Dept. of Physics and Astronomy, University of Delaware, Newark, DE 19716, USA}

\author{K. Andeen}
\affiliation{Department of Physics, Marquette University, Milwaukee, WI, 53201, USA}

% ...


\collaboration{390}{IceCube Collaboration}

\begin{abstract}

This paper presents the results of a search for neutrinos that are spatially and temporally coincident with 22 unique, non-repeating Fast Radio Bursts (FRBs) and one repeating FRB (FRB121102). FRBs are a rapidly growing class of Galactic and extragalactic astrophysical objects that are considered a potential source of high-energy neutrinos. The IceCube Neutrino Observatory's previous FRB analyses have solely used track events. This search utilizes seven years of IceCube's cascade events which are statistically independent of the track events. This event selection allows probing of a longer range of extended timescales due to the low background rate. No statistically significant clustering of neutrinos was observed. Upper limits are set on the time-integrated neutrino flux emitted by FRBs for a range of extended time-windows. 
\\

\end{abstract}

\section{Introduction} \label{sec:intro}
The IceCube Neutrino Observatory, located at the geographic South Pole, is the largest neutrino detector in the world.  Encompassing a cubic kilometer of instrumented ice, IceCube is comprised of 5,160 digital optical modules (DOMs) situated on 86 read-out and support cables or ``strings'' to detect the Cherenkov radiation from charged particles created by neutrino interactions in the Antarctic ice \citep{DOM_MB, detector}. IceCube's instrumentation density is optimized for the detection of neutrinos with energies from 100 GeV -- 10 PeV.  It contains a higher density sub-volume, enabling the detection of neutrinos down to 10 GeV \citep{Deepcore}. IceCube has observed a diffuse flux of high-energy astrophysical neutrinos \citep{2013, HESE3yr, 2yrNorthernTracks}. A study of data collected between April 6, 2008 and July 10, 2018 revealed a 3.3$\sigma$ inconsistency with background expectations for four sources that include NGC 1068 and TXS 0506+056 \citep{PhysRevLett}. In 2022, IceCube found further evidence of neutrino emission from NGC 1068 at a significance of 4.2$\sigma$ \citep{NGC1068}. Despite this growing evidence, the origin of the majority of the diffuse astrophysical flux remains unexplained. In this paper, we present a search for time-dependent neutrino emission from various Fast Radio Bursts (FRBs) using seven years of IceCube's cascade events and set upper limits on the associated neutrino flux for flares of varying duration. 

Transient astrophysical objects are among the primary candidates for producing the astrophysical neutrino flux \citep{Murase_2019}. The first evidence of neutrino emission from a flaring object came from the blazar TXS 0506+056 \citep{TXS2017, TXS}. FRBs are a class of transient astrophysical objects that could contribute to the diffuse neutrino flux \citep{Metzger_2020}. FRBs are periodic or non-periodic transient radio bursts that have Galactic or extragalactic origins. To date, hundreds of FRBs across the entire sky have been detected. Recently, FRB200448 was localized to the Galactic magnetar SGR 1935+2154, suggesting magnetars may be a source of FRBs \citep{chime}. This has been further supported by the polarization of some FRBs \citep{polarized}. While the underlying mechanism that creates FRBs is unknown, it is predicted that this coherent radio emission is the result of an ultra-relativistic shock that propagates into a baryon filled medium. A by-product of this scenario are TeV -- PeV neutrinos that are produced by photo-hadronic interactions on timescales of varying duration after the FRB \citep{Metzger_2020, FRBmagnetars}. Here, we perform a time-dependent stacking search to test if two catalogs of FRBs, one repeating and the other non-repeating, are producing a statistically significant number of neutrinos.

\section{Search for Correlated Cascade Events} \label{methods}
The majority of events detected by IceCube can be divided into two topological classes: cascades and tracks. Cascade signatures are produced by charged-current electron neutrino and tau neutrino interactions as well as all-flavor neutral-current interactions. These interactions typically produce electromagnetic and hadronic showers that have a range of 20 meters or less; this range scales with energy \citep{2014}. Due to the average DOM spacing and light scattering within the ice, these particle showers lead to cascade events having angular resolutions $\sim$10$^\circ$ -- 15$^\circ$ \citep{MESC}. In contrast, track events are the result of muons that are produced by charged-current muon neutrino interactions and have an angular resolution of less than 1$^\circ$ at TeV -- PeV energies \citep{10yr_Tracks}. Track events are subject to larger background rates that stem from the atmospheric muons. To reduce this, track events below a certain energy threshold are filtered out \citep{10yr_Tracks}. The distinctive topology of cascade events allows our selections to include events that have energies down to hundreds of GeV \citep{MESC}. 

To search for correlations between FRBs and cascade events, the energy, spatial, and temporal information of each cascade event is used with the spatial and temporal information of each FRB. This method is similar to that of IceCube's previous FRB analyses, referred to as Six-year Southern Tracks \citep{SixYearSouth} and All-Sky Tracks \citep{AllSkyTracks}, which analyzed track data for correlations.
\\
\subsection{Dataset and FRB Catalogs}

This analysis uses IceCube's Medium Energy Starting Cascade (MESC) data-set. A starting event is classified as an event that contains the neutrino interaction vertex in the detector. This data-set was originally developed to search for all-sky signatures of potential neutrino sources such as the Galactic plane \citep{MESC}. It contains 1,980 cascade events detected from May 2010 to May 2017. This corresponds to 2428 days of IceCube livetime. The first year of data was taken with a 79-string detector configuration and the remaining six years use the complete 86-string configuration. The events in this data-set have energies that range between 270 GeV and 1.6 PeV. 

\begin{figure}[h!]
    \centering
    \includegraphics[width=.95\linewidth]{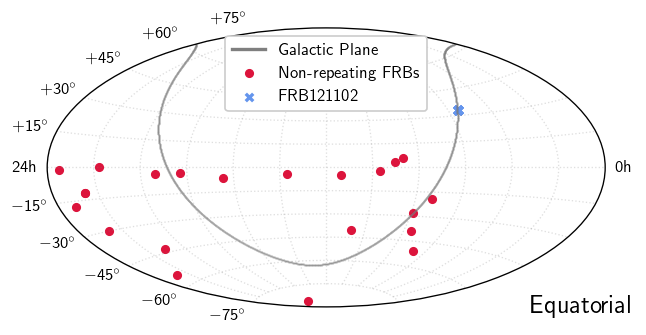}
    \caption{The spatial distribution of FRBs in equatorial coordinates. The 22 unique, non-repeating FRBs are shown in red and FRB121102 is shown in blue. FRB121102 is an extragalactic source located at a declination of 33.15$^{\circ}$ \citep{Spitler_2014}. }
    \label{fig:skymap}
\end{figure}

This analysis focuses on 48 unique bursts that were detected between May 2010 - May 2017; the bursts are separated into two catalogs \citep{FRBcatalog}. As shown in Fig. \ref{fig:skymap}, the first catalog contains 22 non-repeating FRBs that are located at a declination of six degrees or below. The second catalog contains one repeating FRB (FRB121102) that produced 26 bursts within the livetime of the MESC data-set. Since FRBs are an observational class of astrophysical phenomena, non-repeating and repeating FRBs could be associated with different classes of astrophysical objects. As the underlying physical processes may differ, we treat them in this analysis as independent phenomena. These catalogs are a subset of the FRBs that were analyzed in the Six Year Southern Tracks and All-Sky Tracks analyses. This is because the MESC dataset has a livetime period that does not fully overlap with the track datasets. The catalogs of non-repeating and repeating FRBs are independently analyzed through two time-dependent stacking analyses. \\

\begin{figure*}[t]
    \includegraphics[width=.5\linewidth]{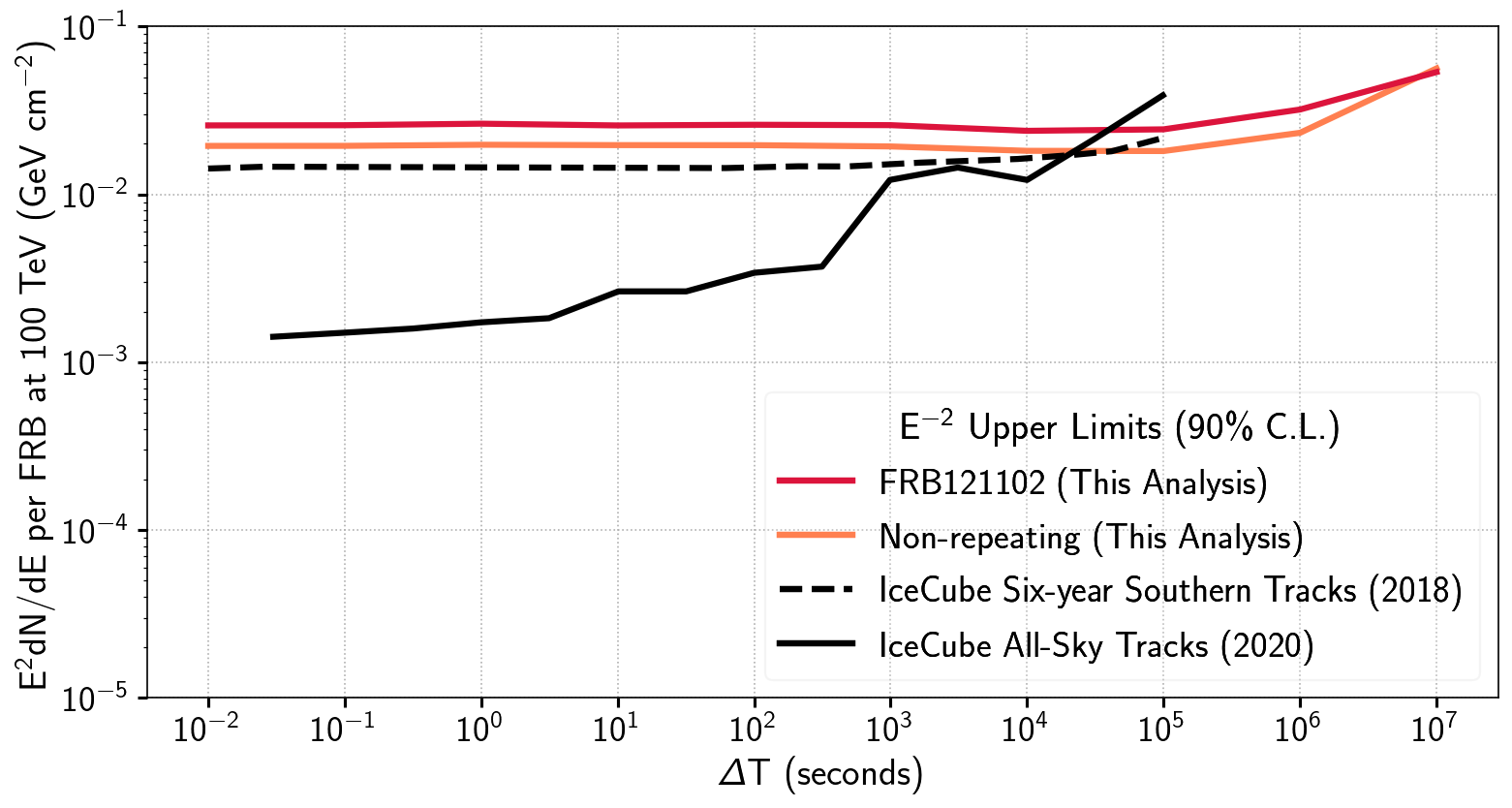}
    \includegraphics[width=.5\linewidth]{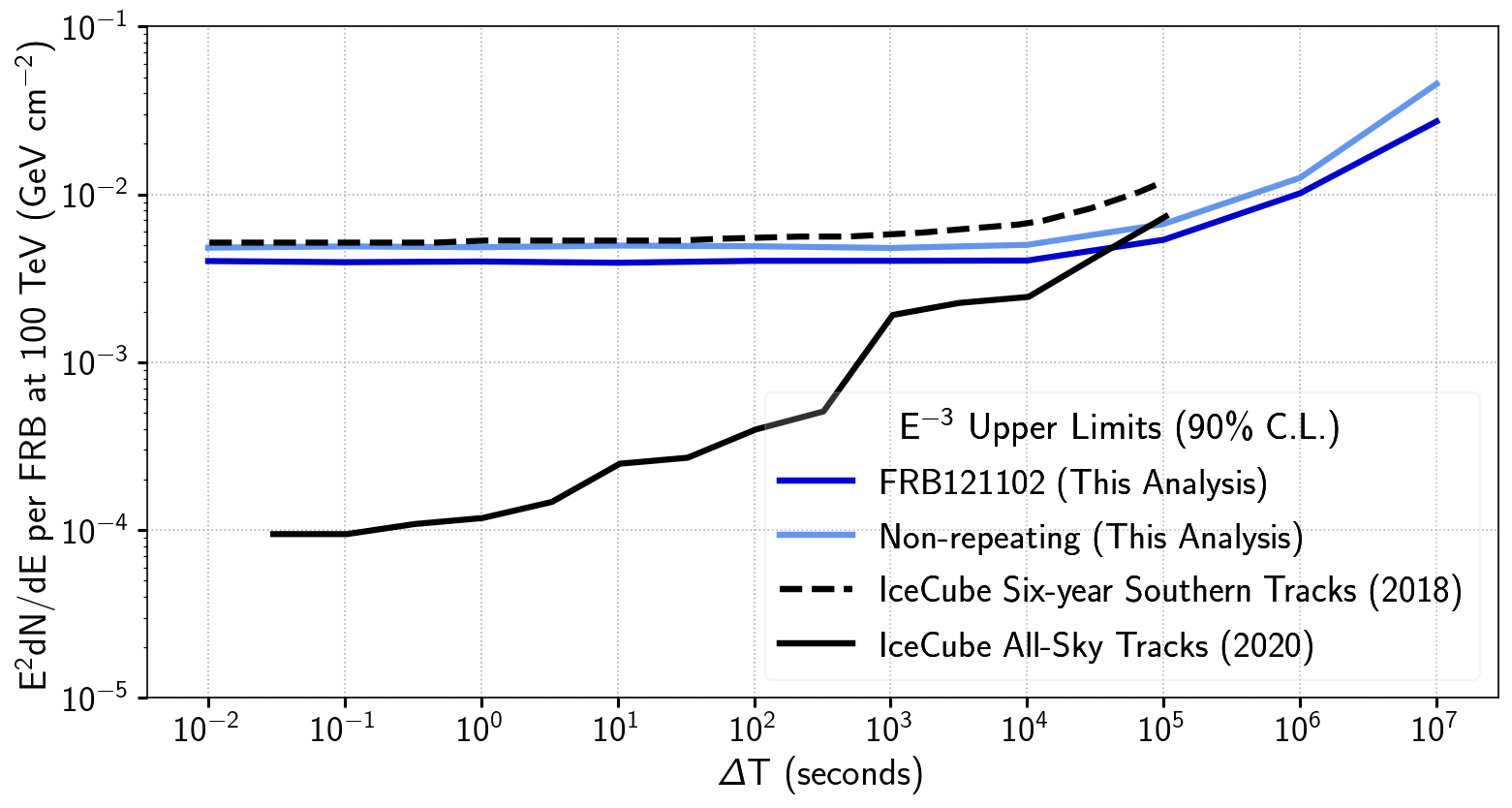}
    \caption{The 90\% confidence level upper limits as a function of time window duration. We assume power law spectra of \textrm{E$^{-2}$} (\textit{left}) and \textrm{E$^{-3}$} (\textit{right}) for the time-integrated neutrino fluxes. The Six-Year Southern Tracks and All-Sky Tracks analyses used statistically independent events; they are included in addition to our limits. Note that previous IceCube analyses used symmetric time windows that are centered on the FRB to search for neutrino emission; this analysis only searches for neutrino emission after the FRB has occurred. Hence \begin{math}\Delta T\end{math}is offset by $\frac{\Delta T}{2}$ between this analysis and IceCube's previous analyses and the time windows do not completely overlap.  }
    \label{fig:upperlimits}
\end{figure*}

\subsection{Analysis Methods}

This analysis uses an unbinned extended maximum likelihood method, similar to the previous analyses \citep{Braun_2010, SixYearSouth}. The likelihood is comprised of spatial, temporal, and energy probability density functions (PDFs). These PDFs characterize how signal-like or background-like a given cascade is. To test if a cascade and FRB are temporally correlated, a search time window, \begin{math} \Delta T, \end{math} is constructed in the interval \begin{math}[t_{FRB},\:t_{FRB} + \Delta T ]\end{math} where \begin{math}t_{FRB}\end{math} denotes when the FRB was detected. Motivated by \cite{Metzger_2020}, we test eight different \begin{math} \Delta T \end{math}that are logarithmically spaced from [10$^{-2}$, 10$^{7}$] seconds after the FRB has occurred.\footnote{The Six Year Southern Tracks and All-Sky Track analyses searched for neutrino emission in the following intervals \begin{math}[t_{FRB}\end{math} - $\frac{\Delta T}{2}$, \begin{math}t_{FRB}\end{math} + $\frac{\Delta T}{2}$].} The spatial PDF accounts for the angular distance between a given FRB and the reconstructed cascade direction. A 2-D Gaussian spatial PDF is used for cascades with an angular resolution less than 7$^\circ$, while for a resolution greater than 7$^\circ$ a Kent Distribution is used. At larger angular resolutions, a Kent Distribution characterizes a probability distribution in curved space and cannot be approximated by a Gaussian distribution in flat space. The energy PDF characterizes the neutrino flux in terms of a powerlaw, \begin{math}E^{-\gamma}\end{math}, where $\gamma$ is the spectral index. We assume a single power-law for the energy flux. From the likelihood, we construct a test statistic (TS) that encompasses our null and alternative hypotheses, 
\begin{equation} \label{ts}
    \textrm{TS} = -2 \log \frac{\mathcal{L}(n_s=0)}{\mathcal{L}(n_s, \gamma)}.
\end{equation}

\noindent The null hypothesis represents background expectations \begin{math}(n_s = 0)\end{math}whereas the alternative hypothesis fits for two parameters: the number of signal events, \begin{math}n_{s},\end{math}and \textrm{$\gamma$}. The number of events in a given time window can vary according to Poisson statistics. In each \begin{math}\Delta T,\end{math}a small number of events are required to observe a statistically significant correlation. To perform a stacking search, the TS in Eq. \ref{ts} is additionally summed over each FRB in the catalog. 

\section{Results} \label{results}

We observed no significant emission in this analysis resulting in \textrm{TS = 0} for every time-window. Upper limits are calculated for both catalogs at the 90\% confidence level for the time-integrated flux per FRB for every \begin{math}\Delta T\end{math} (Figure \ref{fig:upperlimits}). The upper limits assume a flavor ratio ($\nu_{e}$:$\nu_{\mu}$:$\nu_{\tau}$) of 1:1:1 with equal parts from $\nu$ and $\overline{\nu}$. From 10$^{-2}$ to approximately 10$^{6}$ seconds, the upper limits are relatively constant due to a low background rate. The background saturation occurs 10$^6$ seconds after the FRB. 

For \begin{math}E^{-2}\end{math}(Figure \ref{fig:upperlimits} - \textit{left}) and \begin{math}E^{-3}\end{math}(Figure \ref{fig:upperlimits} - \textit{right}) spectra, we find that the stacked upper limits with cascade events are comparable to Six Year Southern Tracks and All-Sky Tracks results (Figure \ref{fig:upperlimits}). For \begin{math}E^{-3},\end{math}this analysis offers improved upper limits with respect to the Six-year Tracks. For \begin{math}E^{-2}\end{math}this analysis is not quite as sensitive as the Six-Year Tracks when searching for neutrino emission in \begin{math}\Delta T\end{math}less than 10$^4$ seconds. The flux allowed by the All-sky tracks analysis increases by an order of magnitude between \begin{math}\Delta T\end{math}that have durations of sub-second to days. This is due to the rapidly increasing background rate which requires more signal events to see a statistically significant correlation. As noted above, this is in contrast with cascade events. Finally, we find that for harder spectra, the catalog of non-repeating FRBs offers slightly more constraining upper limits when compared with FRB121102. This is in contrast with our upper limits for soft spectra where FRB121102 has slightly stronger upper limits.

\begin{figure}[h]
    \centering
    \includegraphics[width=1.0\linewidth]{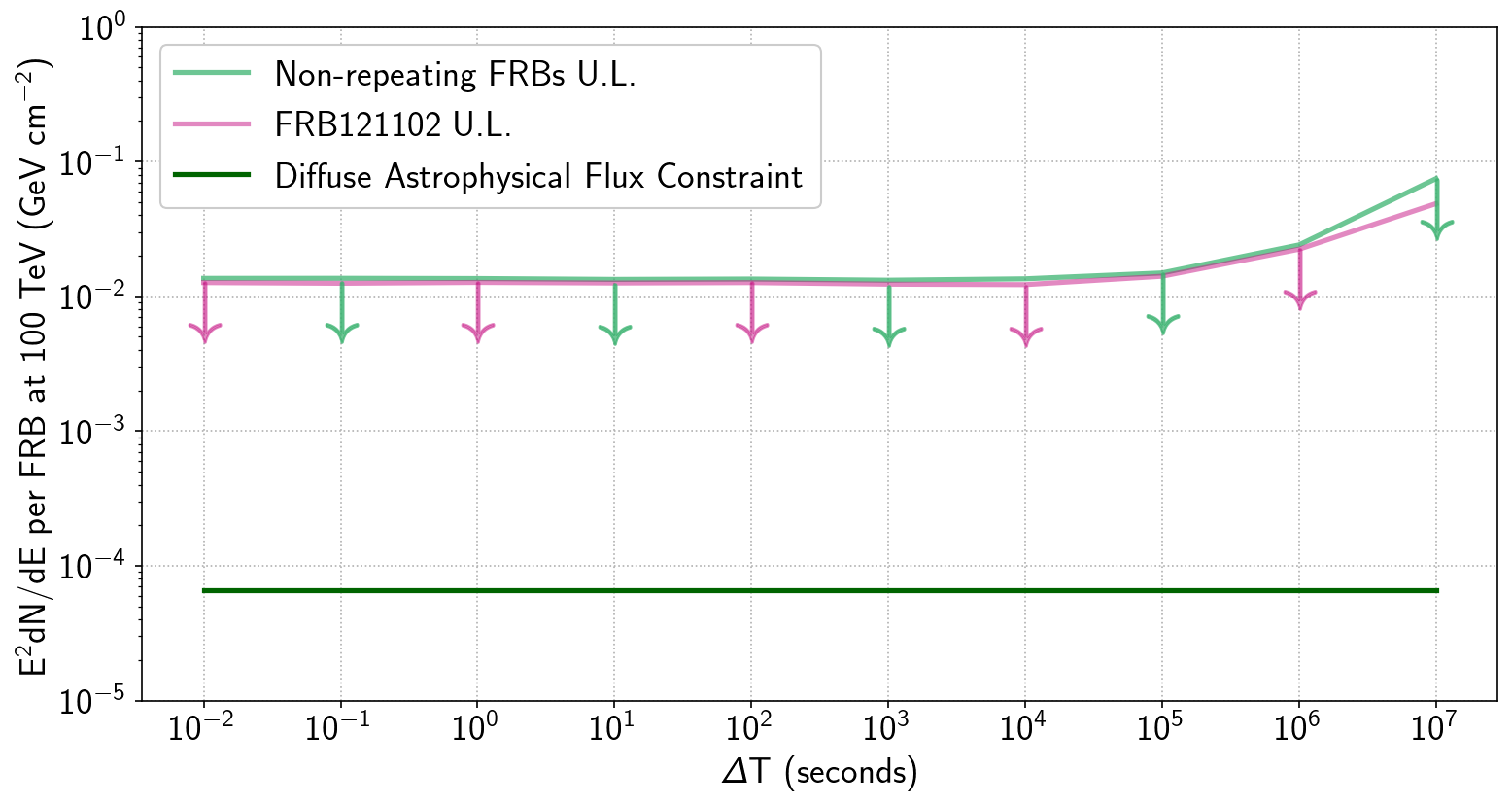}
    \caption{The 90\% confidence level upper limits as a function of time windows assuming the best-fit diffuse spectrum \begin{math}(E^{-2.53})\end{math}measured in cascades \citep{diffusecascade}. The Diffuse Astrophysical Flux Constraint assumes that FRBs are solely responsible for the diffuse neutrino flux. The constraint is calculated by dividing the entire diffuse astrophysical flux equally amongst 820 homogeneous FRBs. Note that we assume 820 FRBs per day from the CHIME experiment's estimations of the FRB all-sky rate \citep{FRB_rate}. }
    \label{fig:diffuse}
\end{figure}

These results are extended to \begin{math}E^{-2.53}\end{math}to draw comparisons to the measured diffuse flux that has been observed in IceCube's cascade analyses \citep{diffusecascade}. The diffuse flux is an all-sky high-energy neutrino flux. Since FRBs are an all-sky phenomena, these results are relevant. Figure \ref{fig:diffuse} compares the per-burst upper limits for each catalog to the diffuse cascade astrophysical flux by integrating the diffuse flux over both the duration of each \begin{math} \Delta T \end{math}and the entire sky then dividing by the Canadian Hydrogen Intensity Mapping Experiment's (CHIME) estimated FRB all-sky rate of 820 FRBs per day \citep{FRB_rate}. This assumes that each burst contributes equally to the diffuse all-sky cascades. This flux per FRB is then compared to the \begin{math}E^{-2.53}\end{math} upper limits from this analysis. Using this method of comparison, we see that the measured diffuse cascade flux establishes the most stringent limit on neutrino emission from FRBs.

\section{Conclusion and Future Outlook} \label{sec:conclusion}
\begin{figure}[h]
    \centering
    \includegraphics[width=1.0\linewidth, trim=4 4 4 4,clip]{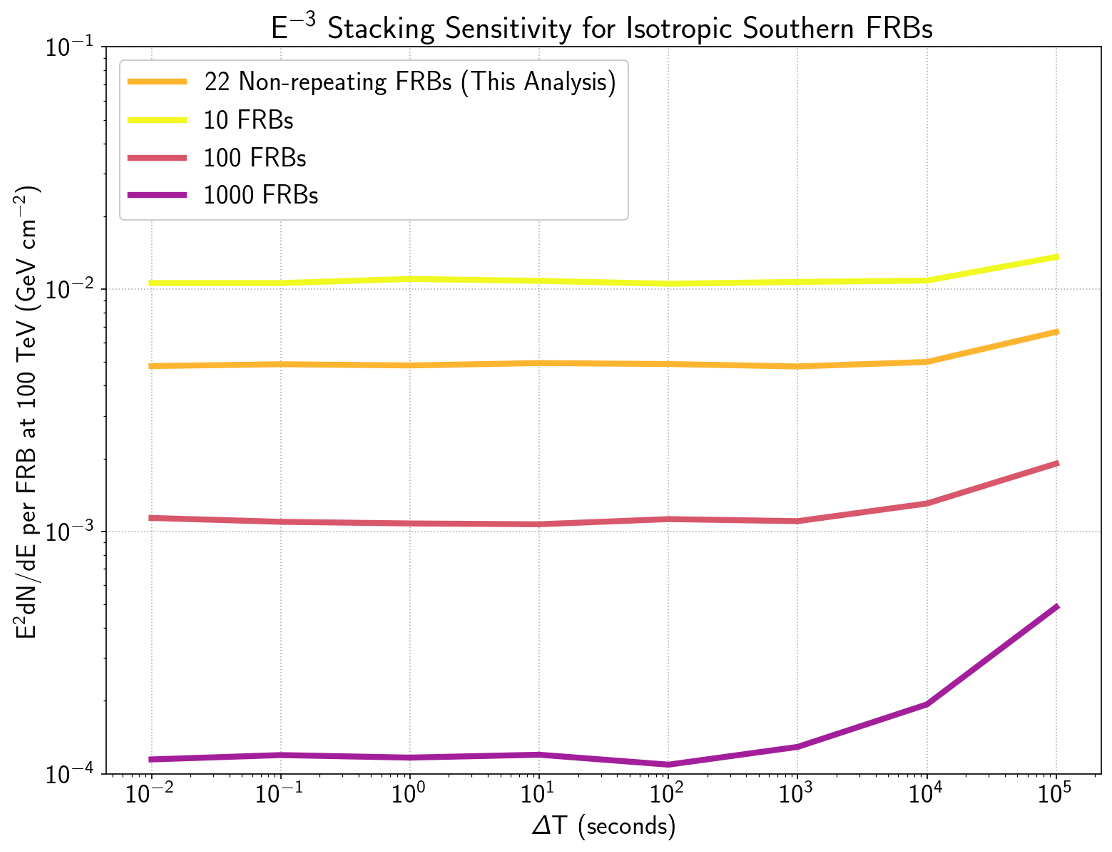}
    \caption{Projected sensitivities for various time-windows durations as the number of FRBs in a catalog increases. We assume a powerlaw spectrum of \begin{math}E^{-3}\end{math}for the time-integrated neutrino fluxes. Future catalogs are expected to have orders of magnitude more FRBs. The projected sensitivities are calculated at the 90\% confidence level and use simulated FRBs that are uniformly distributed in the southern sky.}
    \label{fig:projsens}
\end{figure}

In two independent searches for neutrino emission from 22 unique, non-repeating FRBs and 26 unique bursts from FRB121102, no significant correlation was found. We provide upper limits on the time integrated neutrino flux at the 90\% confidence level for various spectral indices, as shown in Figures \ref{fig:upperlimits} and \ref{fig:diffuse}.  We also provide estimates on the neutrino flux for FRB catalogs of various sizes (Figure \ref{fig:projsens}). These estimates show that for non-periodic FRBs that are isotropically distributed throughout the southern sky, the neutrino flux per FRB decreases as the catalog size increases. In recent years more FRB observatories, such as CHIME, have come online. This has increased the detection rate of both repeating and non-repeating FRBs. However, we were only able to analyze a subset of FRBs that overlapped with our data-set. New data-sets are in preparation that would overlap with 100 -- 1,000 FRBs and potentially offer more stringent upper limits \citep{FRBcatalog, FRB_rate}. 

Overall, this analysis shows that cascades offer sensitive upper limits when performing transient stacking searches. Given that cascades have a low background rate and are considered independent of track events, IceCube's future analyses would benefit from combining track and cascade events when searching for time-dependent neutrino emission from FRBs. This approach can be extended to IceCube's potential searches that aim to observe transient neutrino emission in realtime as well as gamma ray bursts \citep{realtime, GRB}. In addition, the next-generation of IceCube, IceCube-Gen2, will provide the opportunity to conduct more sensitive searches for neutrino emission from transient sources \citep{gen2}. In turn, this will allow us to uncover the origin of high-energy astrophysical neutrinos.

\newpage

\section*{Acknowledgements}
The IceCube collaboration acknowledges the significant contributions to this manuscript from Michael Kovacevich.
USA {\textendash} U.S. National Science Foundation-Office of Polar Programs,
U.S. National Science Foundation-Physics Division,
U.S. National Science Foundation-EPSCoR,
Wisconsin Alumni Research Foundation,
Center for High Throughput Computing (CHTC) at the University of Wisconsin{\textendash}Madison,
Open Science Grid (OSG),
Advanced Cyberinfrastructure Coordination Ecosystem: Services {\&} Support (ACCESS),
Frontera computing project at the Texas Advanced Computing Center,
U.S. Department of Energy-National Energy Research Scientific Computing Center,
Particle astrophysics research computing center at the University of Maryland,
Institute for Cyber-Enabled Research at Michigan State University,
and Astroparticle physics computational facility at Marquette University;
Belgium {\textendash} Funds for Scientific Research (FRS-FNRS and FWO),
FWO Odysseus and Big Science programmes,
and Belgian Federal Science Policy Office (Belspo);
Germany {\textendash} Bundesministerium f{\"u}r Bildung und Forschung (BMBF),
Deutsche Forschungsgemeinschaft (DFG),
Helmholtz Alliance for Astroparticle Physics (HAP),
Initiative and Networking Fund of the Helmholtz Association,
Deutsches Elektronen Synchrotron (DESY),
and High Performance Computing cluster of the RWTH Aachen;
Sweden {\textendash} Swedish Research Council,
Swedish Polar Research Secretariat,
Swedish National Infrastructure for Computing (SNIC),
and Knut and Alice Wallenberg Foundation;
European Union {\textendash} EGI Advanced Computing for research;
Australia {\textendash} Australian Research Council;
Canada {\textendash} Natural Sciences and Engineering Research Council of Canada,
Calcul Qu{\'e}bec, Compute Ontario, Canada Foundation for Innovation, WestGrid, and Compute Canada;
Denmark {\textendash} Villum Fonden, Carlsberg Foundation, and European Commission;
New Zealand {\textendash} Marsden Fund;
Japan {\textendash} Japan Society for Promotion of Science (JSPS)
and Institute for Global Prominent Research (IGPR) of Chiba University;
Korea {\textendash} National Research Foundation of Korea (NRF);
Switzerland {\textendash} Swiss National Science Foundation (SNSF);
United Kingdom {\textendash} Department of Physics, University of Oxford.
\newpage
\bibliography{v5}{}
\bibliographystyle{aasjournal}

\end{document}